\documentclass[aps,twocolumn,rmp,11pt,tightenlines,epsfig]{revtex4}

\usepackage{epsf}
\usepackage{amsfonts}
\usepackage{amssymb}
\usepackage{graphicx}
\usepackage{color}
\usepackage{amsmath}
\usepackage[bookmarksnumbered, bookmarks, breaklinks, linktocpage]{hyperref}

\newcommand{\Tr}        {\mathrm{Tr}}

\newcommand{\ket}[1]    {| #1 \rangle}
\newcommand{\bk}[2]     {\langle #1 | #2 \rangle}
\newcommand{\kb}[2]     {| #1 \rangle \! \langle #2 |}

\newcommand{\cS}        {{\mathcal S}}
\newcommand{\cA}        {{\mathcal A}}
\newcommand{\cE}        {{\mathcal E}}

\newcommand\cF{{\mathcal F}}
\newcommand\hocom[1]{}

\newcommand{\ba}{\begin{eqnarray}}
\newcommand{\ea}{\end{eqnarray}}
\newcommand{\bmath}{\begin{mathletters}}
\newcommand{\emath}{\end{mathletters}}
\newcommand{\ban}{\begin{eqnarray*}}
\newcommand{\ean}{\end{eqnarray*}}

\usepackage{tikz}
\usetikzlibrary{arrows,shapes}

\begin{document}

\tikzstyle{every picture}+=[remember picture]

\title{Emergence of the Classical from within the Quantum Universe}

\author{Wojciech Hubert Zurek}

\address{Theory Division, MS B213, LANL
    Los Alamos, NM, 87545, U.S.A.}

\date{\today}

\begin{abstract} 
Decoherence shows how the openness of quantum systems -- interaction with their environment -- suppresses flagrant manifestations of quantumness. Einselection accounts for the emergence of preferred quasi-classical {\it pointer states}. Quantum Darwinism goes beyond decoherence. It posits that the information acquired by the monitoring environment responsible for decoherence is disseminated, in many copies, in the environment, and thus becomes accessible to observers. This indirect nature of the acquisition of information by observers who use the environment as a communication channel is the mechanism through which objective classical reality emerges from the quantum  substrate: States of the systems of interest are not subjected to direct measurements (hence, not perturbed) by the agents acquiring information about them. Thus, they can exist unaffected by the information gained by observers. 

\end{abstract}
\maketitle

I met Dieter Zeh for the first time at a mini-conference organized by John Wheeler in Austin, Texas, circa 1983. I was then at Caltech, and I was curious about Zeh. I found out about him through Eugene Wigner who visited Wheeler in Texas around the time I submitted the manuscript of my ``pointer basis'' paper to {\it Physical Review}. 

My inspiration for the pointer basis came from the papers on non-demolition measurements involved in gravity wave detection (see e.g., Caves et al, 1980), but Wigner sensed affinity with Zeh's views, and suggested I send him my manuscript. Dieter sent back a copy of (Zeh, 1971) with a note ``look at page 268'' scribbled in pencil (I still have it). I read it (p. 268 starts with a discussion of the loss of the interference pattern due to scattered radiation based on the discussion of the double slit experiment in Feynman's  ``Lectures'') and I also read his 1970 {\it Foundations of Physics} paper. What came through was (in addition to Zeh's allegiance to Everett) conviction that macroscopic systems are difficult to isolate, and so Schr\"odinger equation is not valid. 
I cited Zeh's Fermi School lecture and his {\it Foundations of Physics} paper (Zeh, 1970) in my early work. (The citation to (Zeh, 1971) was added ``in proofs'' to my 1981 paper that introduced pointer basis.) 

I remember that on our first meeting Zeh noted, jokingly, that if we ever wrote a paper together, he would be, alphabetically, the first author. I was always sorry we never did, but I also have a feeling this might have been a frustrating experience for both of us: Dieter was a committed Everettian, and I was resolved to remain open minded about the interpretation. I felt Copenhagen and Everett's interpretations were both incomplete, and more calculating and thinking unconstrained by the allegiance to either was the best way to make progress. 

Dieter drew attention to the fact that macroscopic systems can never be isolated from the environment, and realized that, therefore, one should not expect them to follow unitary dynamics generated by Schr\"odinger equation. This was a valuable insight. 

Given Zeh's adherence to Everett's views (obvious in his earliest papers), I was surprised he did not address the gaping hole in Everett's interpretation -- the selection of the preferred states (or the origin of the ``branches''). He mentioned it on occasion (Zeh, 1973), and even realized that stability should be a criterion. However, in the end he seemed content with the eigenstates of the density matrix as candidates for classicality, and it is clear (and was eventually also accepted by Dieter) that they are unsuitable. 

I have however recently re-read Everett's writings, and I am now less surprised -- this issue is never pointed out in Everett's paper or in his ``long thesis'' (Everett, 1957a,b). Indeed, it appears Zeh (or other Everettians, e.g. DeWitt, 1971) did not fully appreciate its importance. This point is, however, very much appreciated by the present-day leading followers of Everett (see e.g. Kuypers and Deutsch, 2020; Saunders et al, 2010; Wallace, 2012a,b, and references therein).

Dieter and I have met, over the years, several more times. He was at the meeting in Santa Fe (Zeh, 1990). He also contributed (Zeh, 1994) to the proceedings of the meeting in Spain (Halliwell et al, 1994) I helped organize. We have both lectured at the Poincar\'e seminar in 2005 in Paris (for proceedings see Raimond and Rivasseau, 2006). Dieter talked there about ``roots and fruits of decoherence'' (Zeh, 2006). 

The encounters I remember most fondly happened when I spent half a year, month at a time, in Heidelberg under the auspices of the Humboldt Foundation. I had an office on Philosophenweg, in the same building where Dieter had his office before he retired from the University. I met there his colleagues and some of the co-authors of the collection (Joos et al., 2003). There were lively discussions.

The most pleasant and memorable encounters were the dinners at Zeh's home in Waldhilsbach near Heidelberg. When the weather was good, we dined in his garden. During and especially after dinner the conversation turned to physics. There was always good wine, and when Dieter refilled my glass, he usually emphasized (with a twinkle in his eyes) that the chances of encountering police that late at night when I drove back to Heidelberg were minimal.

I think we have both consciously stayed away from the issues of interpretation. Dieter was an enthusiastic Everettian, and I thought that both of the prevailing interpretations were really unfinished projects. Decoherence was the first (and excellent) step rather than the final word on the subject of the quantum-to-classical transition, but I saw it as an interpretation-neutral development, and more a likely beginning of more insights rather than the end. He obviously disagreed, at least about the interpretation-neutral development part.

We have nevertheless agreed about the importance of decoherence for interpretation, but, again, I thought more could be done and calculated. I was, in particular, convinced that the information theoretic tools can be useful in his task, while he seemed skeptical, and thought that information is often brought into quantum foundational discussion as a {\it deus ex machina}.  One subject on which we emphatically agreed (and where information mattered) was the connection between decoherence and the second law in quantum chaotic systems where it yields the entropy production at the Boltzmann-Kolmogorov-Sinai rate (Zurek and Paz, 1994; Zurek, 1998; Zeh, 2007). 

Quantum Darwinism was too early in its development for me to seriously engage Zeh on this subject. I have a feeling he would have been initially skeptical (this was generally the pattern, even with pointer states and einselection), but I also think he might have welcomed it as he eventually welcomed pointer states, even though -- long time after they were defined (Zurek, 1981; 1982) -- he was still favoring the eigenstates of the reduced density matrix as preferred basis  (see Zeh, 1990). He has also, over time, warmed up to the envariant derivation of Born’s rule (Zurek, 2003a; 2005), probably in part because it fulfilled one of the original goals of Everett -- derivation of probabilities that fit well within the the relative state approach -- even though he was also happy to accept Born's rule as an additional postulate.

Part of the problem in discussing these matters with Dieter was also the difference in the attitude we had towards the interpretation of quantum theory. I think he was focused on showing how decoherence turns superpositions into mixtures, and believed that was enough. I think we both agreed that this did not solve all of the ``measurement problem'', but Dieter was happy with the post-decoherence {\it status quo}, and (armed with the Everettian point of view)  felt no need to do more. I -- on the other hand -- had a feeling that the process responsible for the acquisition of information is crucial in how the classical world we perceive emerges from the quantum substrate, and there is more to understand.

Pointer states and environment induced superselection were the first step. Quantum Darwinism is a further  important step in this direction. It clarifies issues left open by decoherence and affirms the importance of  einselection and of the pointer states. 

I mentioned quantum Darwinism briefly in one of our encounters. As expected, there was interest but also resistance. I was therefore surprised when -- reviewing Zeh's papers while preparing this contribution -- I discovered that the key idea of quantum Darwinism is acknowledged (however briefly) in his most recent writings (Zeh, 2016).

I think of this paper as in a way an (obviously belated) attempt to appraise Dieter of the progress on quantum Darwinism and explain my view of its role in the interpretation of quantum theory (including its role in the perception of the ``collapse''). 

\section{Introduction}

Quantum Darwinism (Ollivier et al., 2004; 2005; Blume-Kohout and Zurek, 2005; 2006; 2008; Zurek, 2009; 2014; 2018; Touil et al., 2021) builds on decoherence theory (Joos et al, 2003; Zurek, 2003b, Schlosshauer, 2004; 2007; 2019). It shows how perception of classical reality arises via selective amplification and spreading of information in our fundamentally quantum Universe. Quantum Darwinism (QD) depends on the information transfer from the system to the environment, as does decoherence. However, QD goes beyond decoherence as it recognizes that many copies of the system's pointer states are imprinted on the environment. Agents acquire data indirectly, intercepting environment fragments (rather than measuring systems of interest {\it per se}). Thus, data disseminated through the environment provide us with shared information about stable, effectively classical pointer states. (Humans use primarily photon environment: we see ``objects of interest'' by intercepting tiny fractions of photons that contributed to decoherence.) 

Quantum Darwinism recognizes that the objective classical reality we perceive and we believe in is, in the end, a model constructed by observers whose consciousness relies on indirect means of detecting objects of interest. The confidence we seem to have in this classical reality ignores, to first approximation, 
(or at the very least ignored for a long time) 
the means of its apprehension. One can recall here ``It from Bit''. John Wheeler (1990) emphasized that we construct ``it'', the world ``out there'', from the bits of information (see also “R”, Fig. 7, on p. 195 of Wheeler and Zurek (1983)). 
In the case of quantum Darwinism, this information is delivered to us by the decohering environment.

Indeed, the solid and objective reality we all believe exists is a construct, devised by our consciousness, and based on the second hand information eavesdropped by us from the environment that has also decohered the objects of interest. That environment has two functions: It helps determine the states in which the objects out there can exist, and it delivers to us information about the preferred states of the very objects it has decohered – about their einselected, stable, hence, effectively classical pointer states (Zurek, 1981; 1982). 


Quantum Darwinism is not really ``hypothetical’’ -- it is a ``fact of life''. Its central tenet -- the indirect acquisition of information by observers -- is what actually happens in our quantum Universe. It is surprising that the significance of the indirect means of acquisition of information for the emergence of classical reality was not recognized earlier. Part of the reason may be that decoherence and its role was not identified until about half a century after the interpretation of quantum theory was hotly debated in the wake of the introduction of the Schr\"odinger equation. By then the subject of interpretation was considered settled, and, hence, uninteresting, or at the very least more a domain of philosophy rather than physics. 

But perhaps an even more important reason for the delayed appreciation of the importance of the information transmitted by the decohering environment is that information theory -- a crucial tool in the analysis of information flows and in quantifying redundancy (that is a consequence of amplification which in turn follows as a natural consequence of decoherence) -- was formulated (by Shannon, in 1948) over twenty years after the advent of modern quantum mechanics. As the discussions of the interpretation of quantum theory by Bohr, Heisenberg, Schr\"odinger, and others started immediately after Schr\"odinger's equation was introduced, the interpreters at the time (including von Neumann, Dirac, London and Bauer, and others, in addition to the forefathers mentioned above) did not have the information-theoretic tools needed to take into account what appears to be, in retrospect, evident -- that the information we have about what exists is crucial for what we believe exists. 

It is nevertheless interesting that even these early discussions appealed (at least occasionally) to the ``irreversible act of amplification’’. Quantum Darwinism recognizes that amplification is a natural and an almost inevitable consequence of decoherence, and that it does not need to involve special arrangements present in Geiger counters, photographic plates, or cloud chambers that were invoked in the early discussions. 


The inevitable byproduct of decoherence is then, typically, abundance of the copies of information about the preferred states in the environment. This information is often (but not always) available to the observers. When it is available, fractions of the environment can deliver to agents multiple copies of the data about the pointer states. Agents who receive this indirect information from distinct environment fragments will agree on the objective reality of the states they infer from their data.

Not all decohering environments are equally useful as communication channels. Light excels, and we, humans, rely largely on photons. Other senses can also provide us with useful information. One can -- in all of the examples of perception -- point out redundancies in the information carried by the environments that are employed, although the corresponding communication channels may not be as straightforward to quantify as is the case for light (Riedel and Zurek, 2010; 2011; Zwolak et al., 2014). 

\section{Solving the Measurement Problem}

The literature on the foundations of quantum theory abounds in categorical statements to the effect that an idea (e.g., Everett interpretation or decoherence) does or does not solve the measurement problem. Such claims are only rarely accompanied by an attempt to explain what is the measurement problem. One has a feeling that even presenting its definition would be controversial as there are different reasons to feel discomfort with what happens in quantum measurements, and, more generally, with quantum physics.

Nevertheless, and at the risk of inciting a controversy, I now undertake the task of defining the measurement problem, as we need to describe it before we discuss the extent to which consciousness is implicated. 
As is widely appreciated, the classical ideal of a single unique preexisting state that, for composite systems, has a Cartesian structure (allowing separately for a definite state of the classical system $ \mathsf S$ and the classical apparatus or agent $ \mathsf A$) is inconsistent with the unitarity of quantum evolutions and with the structure of entangled states typically resulting from the interaction of a quantum $\cS$ and $\cA$.

What is then left to do, if this goal of recovering a fundamentally classical world
is out of reach? The obvious answer is the analysis of perceptions. Would they look any different from what we experience in a Universe that is fundamentally quantum? 

We emphasize, in the previous paragraph, the phrase `such as ours’. It is an important caveat, a sign of departure from complete generality and towards explaining perception of reality in {\it our} rather than some generic Universe. A different hypothetical Universe may not have the properties that we take for granted in our Universe and that will matter for the ensuing considerations. Similarly, abilities of the agents should be constrained by the assumptions about their means of perception before we attempt to understand what they perceive. 

We shall make, in particular, three important assumptions. To begin with, we shall assume that, at energies relevant for the measurement we carry out, one can identify systems (such as the system $\cS$ or the apparatus $\cA$) that maintain their identity over timescales relevant for measurements or perceptions. 

Existence of systems is essential in stating the measurement problem, as without systems the state of the whole $\cS \cA$ evolves unitarily and deterministically, and the issue of the definite measurement outcomes does not arise -- there is no need (and no way) to ask about the outcomes when there is just a single state of the whole indivisible $\cS \cA$. It follows that we shall also not apologize for assuming existence of other systems (such as the environment $\cE$, with its subsystems) essential for the discussion of decoherence and quantum Darwinism.

Nevertheless, this assumption raises the question -- how are the systems defined? It was noted earlier (Zurek, 1998b) and is still an open questions. We shall not attempt to address it here. 

For the purpose of our discussion we shall assume that systems exist and that what they are depends to at least some on the past of the part of Universe where they exist (e.g., a chair was made, a cat was born and grew up, et.). We can (as a consequence of quantum Darwinism) relax and make more precise what is expected of systems: Basically, we expect ``systems of interest'' to make imprints of their states on the environment. 

This is a strong assumption, but both photons and air would qualify here as environments, so we make it in recognition of the particular features of the Universe we inhabit. 

The second assumption we shall make is that the Hamiltonians present in the Universe are local. This is certainly consistent with what we know about our Universe. Locality is essential if causality is to be observed, and depends on the existence of space (hence, presumably, time). Locality of interactions does not preclude nonlocality of states of composite quantum systems.

Last not least, and as a third assumption, we assume that the agents that control measuring devices and acquire information are a bit like us. 
To start with the obvious, agents should inhabit -- should be embedded -- in the Universe rather than study its evolution ``from the outside’’. They should be also capable of making measurements using local interactions we have already assumed, so their senses should be somewhat like ours. And they should be able to make records of the outcomes and process information in these records using logical circuits not unlike our neurons or classical computers.

In fact, we shall make an even stronger assumption that the agent's senses work much like ours. That is, we assume that agents utilize excitations in the environments (like we use photons and phonons) to gather information. 

We shall focus on photons as they deliver vast majority of our data, and serve as the principal communication channel through which we perceive our world, including the outcomes of the measurements. Other channels (except perhaps hearing, which relies on sound waves that, because of their wave nature, have properties similar to light) are somewhat different and less relevant for the problem at hand, as taste, smell, or touch rarely arise in discussions of quantum measurements. Nevertheless, the conclusions we shall arrive at can be reasonably easily ({\it mutatis mutandis}, and in presence of the other assumptions, such as the locality of interactions) applied to these other channels. 


These assumptions are adopted to simplify the discussion. They are likely stronger than necessary, and it is an interesting question to see whether they can be relaxed and what would be the consequences of their modifications. This is an intriguing area of research that is only in its infancy but is already leading to interesting insights (see e.g. Brand\~ao, Piani, and Horodecki, 2016; Knott et al., 2018; Qi and Ranard, 2020; Baldij\~ao et al, 2020).

With these preliminaries out of the way, we are now ready to tackle the issue of perception of being conscious of a classical world while embedded within a quantum Universe. We end this section by drawing a sharp distinction between the two similar but, in fact, quite different ways of enquiring about the origin of the ``everyday classical reality''. Thus, one can either ask ``Why do we {\it perceive} our world as classical?'', or ``Why our world {\it is} classical?''.  

We shall pursue the answer to the first of these two questions, understood as the question about the perception of objective classical reality -- effectively classical states and correlations. The second question also has a definite answer -- our Universe is {\it not} fundamentally classical, even on the macroscopic level that includes observers and measuring devices. This point was in effect conceded even by Bohr, when in his famous double slit debate he has defended consistency of quantum mechanics by invoking Heisenberg's indeterminacy for the movable slit in the apparatus proposed by Einstein, even though that slit was an integral part of measuring equipment (hence, according to the Copenhagen interpretation, it should have been {\it ab initio} classical). 

\section{Observers in the Classical and in the Quantum Universe}

In the classical Newtonian setting it was possible to imagine that states are an exclusive property of systems, and that observers perceive them ``as they are'', that they see what really exists. Thus, one could envisage such a fundamental ontic state of a system existing independently of the epistemic state -- of the imprecise information observer has about it. Epistemic states could differ, depending on what and how much observers know about the same fundamental ontic state. For instance, one could distinguish between insiders and outsiders -- observers who knew that state more or less precisely. In classical settings one could also imagine that the only limit on how well an insider knew the fundamental state had to do with the resolution of measurements used to determine it or perhaps with the memory capacity of the observer (as knowing the state more precisely generally involves more data, hence, a lengthier description). Hence, one could think of a fundamental ontic state as a limit approached via a sequence of epistemic states corresponding to increasingly accurate measurements.

In our quantum Universe einselection determines preferred states of the objects of interest, but it does not constrain observers to measure exclusively these preferred states. Thus, one might be concerned that einselection is inconclusive by pointing out that, in principle, agents could measure observables other than the pointer observable. If the aim of measurement is to predict, this would make little sense, but such concerns about the answers offered by decoherence have been raised on occasion (see e.g. S\'anchez-Cañizares, 2019). Quantum Darwinism puts them conclusively to rest.

This danger of disagreement between observers is nevertheless, in view of the preceding remarks about the preferences imposed by the einselection -- a somewhat fanciful motivation. It is in observer's interest to measure pointer observables. So, why do we really need to investigate quantum Darwinism?

The answer is simple: In our Universe environments really act as communication channels. Quantum Darwinism actually happens, and how it happens settles the issue of the alignment of measurements between different observers (and of the emergence of objective classical reality agents can be aware of) much more decisively than decoherence alone. 

Decoherence inevitably implies information acquisition by the environment, so its transmission is a natural consequence.  
Quantum Darwinism recognizes that the environments do not just destroy quantum coherence, but that systems -- while they are being monitored -- imprint multiple copies of information about their preferred pointer states onto the subsystems of that environment. Our senses do not couple objects of our interest directly with our information storage repositories. Rather, we eavesdrop on the information imprinted, in many copies, in the environment. This constrains what we can perceive and be conscious of.

Einselection turns out to be just one of the two functions of the ``environmental monitoring and advertising agency’’. Over and above suppression of the uncomfortably quantum information, monitoring by the environment  produces multiple copies of the data about states that can survive repeated copying -- that is, preferred pointer states. Some (but not all) environments disseminate this information intact, putting it within the reach of agents, who can then get it indirectly (that is, without the danger of disrupting the system with direct measurements).

\subsection{Seeing is Believing}

The requirement of redundancy appears to be built into our senses, and in particular, our eyesight: The wiring of the nerves that pass on the signals from the rods in the eye -- cells that detect light when illumination is marginal, and that appear sensitive to individual photons (Nam et al., 2014) -- tends to dismiss cases when fewer than $\sim 7$ neighboring rods fire simultaneously (Rieke and Baylor, 1998). Thus, while there is evidence that even individual photons can be (occasionally and unreliably) detected by humans (Tinsley et al., 2016), redundancy (more than one photon) is usually needed to pass the signal onto the brain. 

This makes evolutionary sense -- rods can misfire, so such built-in veto threshold suppresses false alarms. Frogs and toads that ``make their living’’ by hunting small, poorly illuminated prey, have apparently lower veto thresholds, possibly because ejecting their tongue at a non-existent target is worth the risk, while missing a meal is not. Moreover, amphibians are cold-blooded, so they may not need to contend with as much noise, as thermal excitation of rods appears to be the main source of ``false positives’’.

Quantum Darwinism relies on repeatability. As observers perceive outcomes of measurements indirectly -- e.g., by looking at the pointer of the apparatus or at the photographic plate that was used in a double-slit experiment -- they will depend, for their perceptions, on redundant copies of photons that are scattered from (or absorbed by) the apparatus pointer or the blackened grains of emulsion. 

Thus, repeatability is not just a convenient assumption of a theorist: This hallmark of quantum Darwinism is built into our senses. And the discreteness of the possible measurement outcomes -- possible perceptions -- follows from the distinguishability of the preferred states that can be redundantly recorded in the environment (Zurek, 2007; 2013).

What we are conscious of is then -- it appears -- based on redundant evidence. Quantum Darwinist update to the existential interpretation is to assert that states exist providing that one can acquire redundant evidence about them indirectly -- from the environment. This of course presumes stability in spite of decoherence (so there is no conflict with the existential interpretation that was originally formulated primarily on the basis of decoherence, Zurek, 1993; 1998b) but the threshold for objective existence is nevertheless raised. 

\subsection{Decoherence of Records and Information Processing}

Decoherence affects record keeping and information processing hardware (and, hence, information retention and processing abilities) of observers. Hence, it is relevant for our consciousness. As was already noted some time ago by Tegmark (2000),  individual neurons decohere on a timescale very short compared to e.g. the ``clock time’’ on which human brain operates or other relevant timescales. Therefore, even if somehow one could initiate our information processing hardware in a superposition of its pointer states (which would open up a possibility of being conscious of superpositions), it would decohere almost instantly. The same argument applies to the present day classical computers. Thus, even if information that is explicitly quantum (that is, involves superpositions or entanglement) was inscribed in computer memory, it would decohere and (at best) become classical. (More likely, it would become random nonsense.)

It is a separate and intriguing question whether a robot equipped with a quantum computer could ``do better’’ and perceive quantumness we are bound to miss. 
However, if such a robot relied (as we do) on the fragments of the environment for the information about the system of interest, it could access only the same information we can access. This information is classical, and quantum information processing capabilities would not help -- only pointer states can be accessed through this communication channel.

Existential interpretation -- as defined originally (e.g., Zurek, 1993) -- relied primarily on decoherence. Decoherence of systems immersed in their environments leads to einselection: Only preferred pointer states of a system are stable, so only they can persist. Moreover, when both systems of interest and agent’s means of perception and information storage are subject to decoherence, only correlations of the pointer states of the measured systems and the corresponding pointer states of agent’s memory that store the outcomes are stable -- only such Cartesian correlations (where each object can have a definite if unknown state) can persist. 

Decoherence, one might say, ``strikes twice’’: It selects preferred states of the systems, thus defining what can persists, hence, exist. It also limits correlations that can persist, so that observer’s memory or the apparatus pointer will only preserve correlations with the einselected states of the measured system when they are recorded in the einselected memory states.

This is really the already familiar discussion of the quantum measurement problem with one additional twist: Not just the apparatus, but also the systems of everyday interest to observers are subject of decoherence. Thus, the post-measurement correlations (when investigated using discord) must now be ``classical-classical’’,
(Piani, Horodecki, Horodecki, 2008), 
while in quantum measurements only the apparatus side was guaranteed to be einselected (hence, certifiably classical).

\section{Quantum Darwinism, Classical Reality, and Objective Existence}

Monitoring of the system by the environment (the process responsible for
decoherence) will typically leave behind multiple copies of its pointer states in $\cE$. Pointer states are favored -- only states that can survive decoherence can produce information theoretic progeny in this manner (Zurek, 2007; 2013). Therefore, only information about
pointer states can be recorded redundantly. States that can survive decoherence
can use the same interactions that are responsible for einselection to proliferate information about
themselves throughout the environment.


Quantum Darwinism (Zurek, 2003b; 2009; 2014) recognizes that observers use the environment as
a communication channel to acquire information about pointer states indirectly, leaving the system of interest untouched
and its state unperturbed. Observers can
find out the state of the system without endangering its existence (which would be inevitable in direct measurements). Indeed, readers of this text are -- at this very moment -- intercepting
a tiny fraction of the photon environment by their eyes to gather information. 

This is
how virtually all of our information is acquired. A direct measurement is not what we do. Rather, we
count on redundancy, and settle for information that exists in many copies. This is how objective existence -- the cornerstone of classical reality -- arises in the quantum world. 

\begin{figure}[tb]
\begin{tabular}{l}
\vspace{-0.15in} 
\includegraphics[width=3.7in]{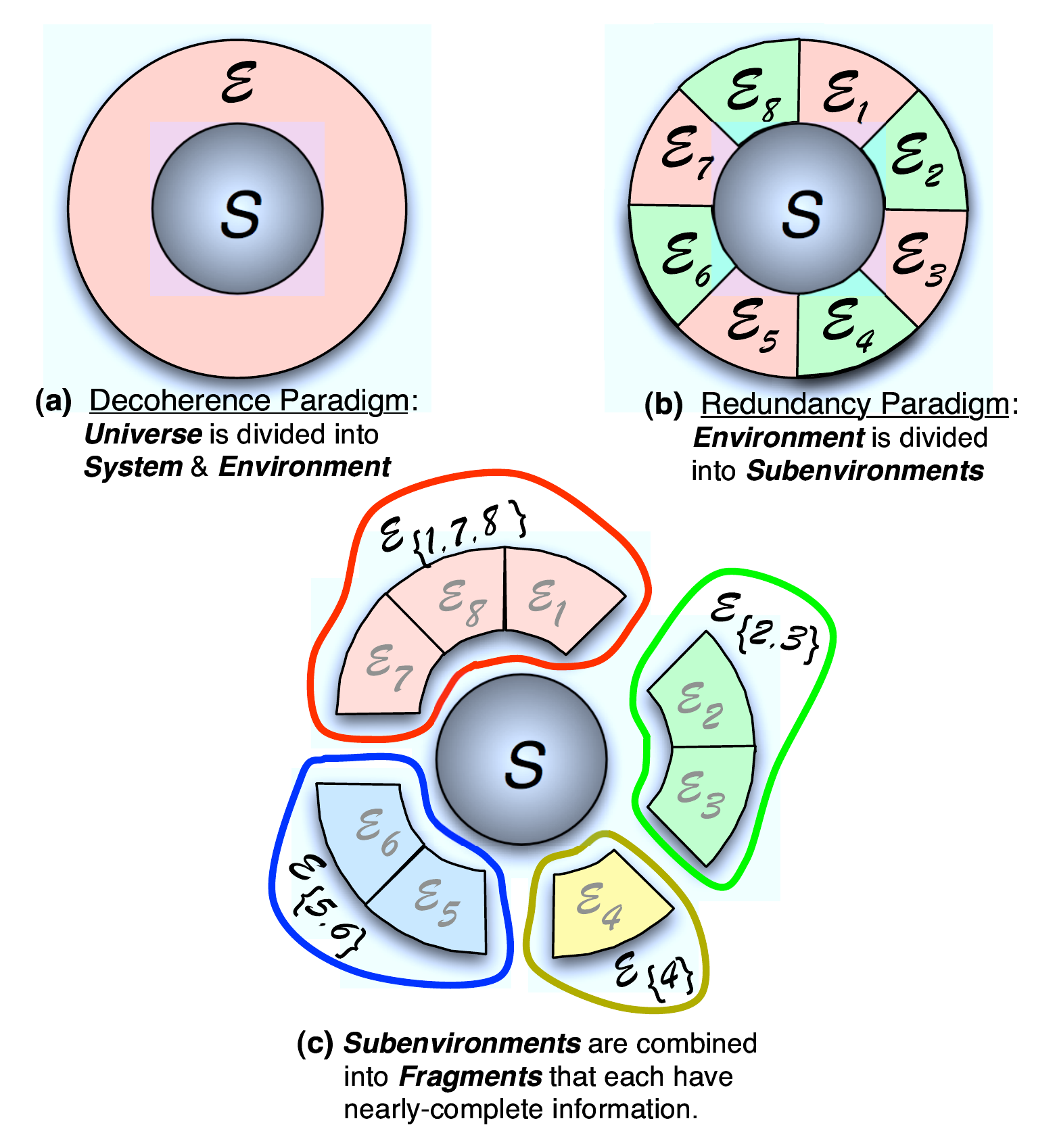}\\
\end{tabular}
\caption{Quantum Darwinism recognizes that environments consist of many subsystems, and that observers acquire information about system of interest $\cS$ by intercepting copies of its pointer states deposited in the fragments $\cal F$ of $\cE$ as a result of decoherence.
}
\label{EnvSubdivision}
\end{figure}

\subsection{Proliferation of Information and Einselection}

Quantum Darwinism was introduced relatively recently. 
Previous studies of the records ``kept'' by the environment were focused on its effect on the state of the system, and not on their utility. Decoherence is a case in point, as are some of the studies of the decoherent histories approach (Gell-Mann and Hartle, 1994; Halliwell 1999).  

The exploration of quantum Darwinism in specific models has started at the beginning of this millennium (Ollivier et al., 2004; 2005; Blume-Kohout and Zurek, 2005; 2006, 2008).
I do not intend to review all of the results of this ongoing research.
The basic conclusion of these studies is, however, that the dynamics responsible for decoherence is
also capable of imprinting multiple copies of the pointer basis on the environment. Moreover, while
decoherence is always implied by quantum Darwinism, the reverse need not be true.  For instance, 
 when the environment is completely mixed, it cannot be used as
a communication channel, but it will still suppress quantum coherence in the system and lead to einselection of pointer states.

Let us consider a simple example of quantum Darwinism: For many subsystems, $\cE=\bigotimes_k \cE^{(k)}$, the initial pre-decoherence state of the system and the environment, $(\alpha \ket \uparrow 
+ \beta \ket \downarrow) 
\ket { {\varepsilon^{(1)}_0} {\varepsilon^{(2)}_0} {\varepsilon^{(3)}_0}...}$ evolves into a ``branching state'';
$$ \ket {\Upsilon_{\cS\cE}} = \alpha \ket \uparrow \ket { {\varepsilon^{(1)}_\uparrow} {\varepsilon^{(2)}_\uparrow}{\varepsilon^{(3)}_\uparrow}... } + \beta \ket \downarrow \ket { {\varepsilon^{(1)}_\downarrow} {\varepsilon^{(2)}_\downarrow}{\varepsilon^{(3)}_\downarrow}...} \eqno(1)$$
 The state $\ket {\Upsilon_{\cS\cE}}$ represents many records inscribed in its fragments, collections of subsystems of $\cE$ (Fig. 1). 
This means that the state of $\cS$ can be found out by many, independently, and indirectly---hence, without disturbing $\cS$. This is how evidence of objective existence arises in our quantum world.

Linearity assures that all the branches persist: collapse to one outcome in Eq. (1) is not in the cards.
However, large $\cE$ can disseminate information about the system.

\hocom{ The existence of redundant copies of pointer states implies that observables which do not commute with
the pointer observable are inaccessible. The simplest model of quantum Darwinism that illustrates this
is a somewhat contrived arrangement of many ($N$) target qubits that constitute subsystems
of the environment interacting via a {\it controlled not} ({\tt c-not}) with a single control qubit $\cS$. As time goes on,
consecutive target qubits become imprinted with the state of the control $\cS$:
$$(a\ket 0+ b\ket 1)\otimes \ket {0_{\varepsilon_1}} \otimes \ket {0_{\varepsilon_2}}\dots \otimes \ket {0_{\varepsilon_N}} \Longrightarrow $$
$$
(a\ket 0\otimes \ket {0_{\varepsilon_1}} \otimes \ket {0_{\varepsilon_2}} + b\ket 1\otimes \ket {1_{\varepsilon_1}} \otimes \ket {1_{\varepsilon_2}})\dots \otimes \ket {0_{\varepsilon_N}} \Longrightarrow
$$
$$
a\ket 0\otimes \ket {0_{\varepsilon_1}} \otimes \dots \otimes \ket {0_{\varepsilon_N}} + b\ket 1\otimes \ket {1_{\varepsilon_1}} \dots \otimes \ket {1_{\varepsilon_N}} \ .  $$
This simple dynamics creates multiple records of the logical basis -- pointer states -- of the system in
the environment. The existence of the preferred pointer basis that is untouched by the interaction is
essential. As we have seen earlier, this is possible -- such quantum jumps emerge from the purely
quantum core postulates (o) - (iii).}

\subsection{Mutual Information in Quantum Correlations}

To develop theory of quantum Darwinism we need to quantify information between fragments of the environment and the system.
Quantum mutual information is a convenient tool we shall use for this purpose.

The mutual information between the system $\cS$ and a fragment $\cF$ (that will play the role of the apparatus $\cA$) can be computed using their von Neumann entropy:
$$H_X=-\Tr \rho_X \lg \rho_X \ . \eqno(2)$$ 
Given the density matrices of the system $\cS$ and a fragment $\cF$ mutual information is simply; 
$$ I(\cS : \cF)=H_\cS+ H_{\cF} - H_{\cS, \cF} \ .
\eqno(3)$$

We already noted the special role of the pointer observable. It is stable and, hence, it leaves behind
information-theoretic progeny -- multiple imprints, copies of the pointer states -- in the environment.
By contrast, complementary observables 
are destroyed by the interaction with a single subsystem of $\cE$. They can in principle still be accessed,
but only if {\it all} of the environment is measured. Indeed, because we are dealing with quantum
systems, things are much worse than that: The environment must be measured in
precisely the right (typically global) basis along with $\cS$ to allow for such a reconstruction. Otherwise, the accumulation of errors over multiple
measurements will lead to an incorrect conclusion and re-prepare the environment, so that it is
no longer a record of the pre-decoherence state of ${\cal S}$, and phase information is irretrievably lost.

\begin{figure}[tb]
\begin{tabular}{l}
\includegraphics[width=3.5in]{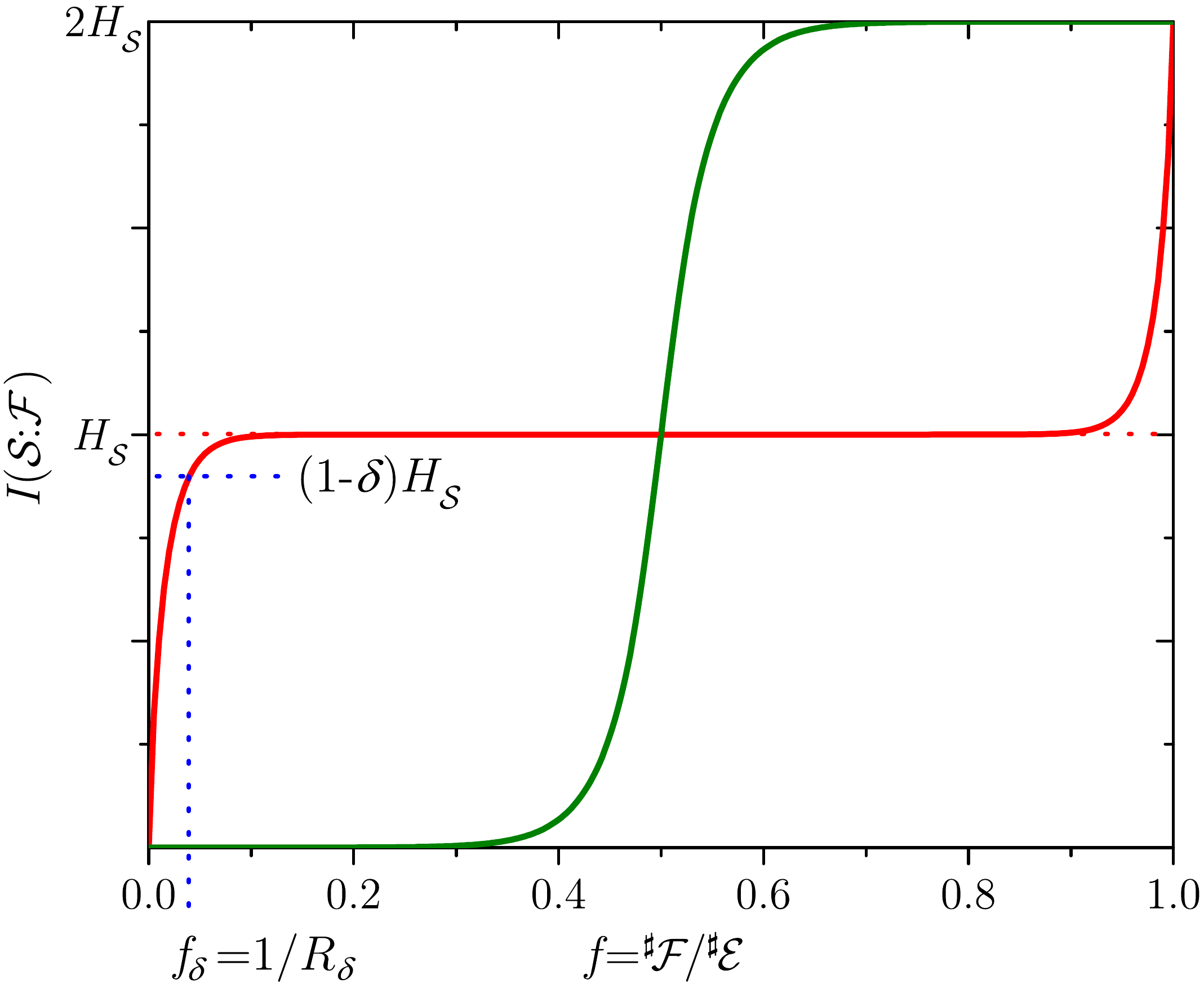}\\
\end{tabular}
\caption{Information about the system contained in a fraction $f$ of the environment. The red plot shows a typical $I(\cS : \cF_f)$ established by decoherence. Rapid rise means that nearly all classically accessible information is revealed by a small fraction of $\cE$. It is followed by a plateau: additional fragments only confirm what is already known. Redundancy ${\cal R_\delta} = 1 / f_\delta$ is the number of such independent fractions. Green plot shows $I(\cS : \cF_f)$ for a random state in the composite system ${\cS\cE}$. In this case, almost no information is revealed until nearly half of the environment is measured.}
\label{RedPIP}
\end{figure}

When each environment qubit is a perfect copy of the state $\cS$, redundancy in this simple example is 
given by the number of fragments -- that is, in this case, by the number of the environment qubits -- that
have complete information about $\cS$. In this simple case there is no reason to
define redundancy in a more sophisticated manner.  Such a need arises in more realistic
cases when the records in individual subsystems of $\cE$ are imperfect.

\subsection{Objective Reality from Redundant Information}

An environment fragment $\cF$ can act as apparatus with a (possibly incomplete) record of $\cS$. When $\cE_{\backslash \cF}$ (`the rest of the environment') is traced out, $\cS\cF$ decoheres, and the reduced density matrix describing joint state of $\cS$ and $\cF$ is:
$$ \rho_{\cS\cF} = \Tr_{\cE \backslash \cF} \kb {\Psi_{\cS\cE}}{\Psi_{\cS\cE}}= $$
\vspace{-0.25in}
$$=|\alpha|^2 \kb \uparrow \uparrow \kb {F_\uparrow} {F_\uparrow} + |\beta|^2 \kb \downarrow \downarrow \kb {F_\downarrow}{F_\downarrow} \eqno(4) $$
When $\bk {F_\uparrow} {F_\downarrow} =0$, $\cF$ contains perfect record of the preferred states of the system. 
In principle, each subsystem of $\cE$ may be enough to reveal its state, but this is unlikely. Typically, one must collect many subsystems of $\cE$ into $\cF$ to find out about $\cS$.

The redundancy of the data about pointer states in $\cE$ determines how many times the same information can be independently extracted---it is a measure of objectivity. 
The key question of quantum Darwinism is then: {\it How many subsystems of $\cE$---what fraction of $\cE$---does one need to find out about $\cS$?}. The answer is provided by the mutual information 
$$I(\cS : \cF_f)=H_\cS + H_{\cF_f} - H_{\cS \cF_f} \ , $$
the information about $\cS$ available from $\cF_f \ , $
that can be obtained from the fraction $f= \frac { \sharp \cF } { \sharp \cE }$ of $\cE$ (where $\sharp \cF$ and $ \sharp \cE$ are the numbers of subsystems). 

In case of perfect correlation a single subsystem of $\cE$ would suffice, as $I(\cS : \cF_f)$ jumps to $H_\cS$ at $f=\frac 1 {\sharp \cE}$. The data in additional subsystems of $\cE$ are then redundant.
Usually, however, larger fragments of $\cE$ are needed to find out enough about $\cS$. Red plot in Fig. 2 illustrates this: $I(\cS : \cF_f)$ still approaches $H_\cS$, but only gradually. The length of this plateau can be measured in units of $f_\delta$, the initial rising portion of $I(\cS : \cF_f)$. It is defined with the help of the {\it information deficit} $\delta$ observers tolerate:
$$ I(\cS : \cF_{f_\delta}) \ge (1- \delta) H_\cS \eqno(5) $$
Redundancy is the number of such records of $\cS$ in $\cE$:
$$ {\cal R_\delta} = 1 / f_\delta \eqno(6)$$
$ {\cal R_\delta}$ sets the upper limit on how many observers can find out the state of $\cS$ from $\cE$ independently and indirectly. In models, and especially in the photon scattering analyzed extending the decoherence model of Joos and Zeh (1985) redundancy $\cal R_\delta$ is huge (Riedel and Zurek, 2010; 2011; Zwolak et al, 2014) and depends on $\delta$ only weakly (logarithmically). 


This is `quantum spam': $\cal R_\delta$ imprints of the pointer states are broadcast throughout the environment. Many observers can access them independently and indirectly, assuring objectivity of pointer states of $\cS$. Repeatability is key: States must survive copying to produce many imprints. 


Insights into the nature of the constraints imposed by unitarity in the process of copying (Zurek, 2007; 2013) show when, in spite of the no-cloning theorem 
repeated transfers of information are possible. Distinguishability of the ``originals'', the states that can be repeatably copied, is essential. Discrete preferred states set the stage for quantum jumps.
Copying yields branches of records inscribed in subsystems of $\cE$. Initial superposition yields superposition of branches, Eq. (1), so there is no literal collapse.
However, fragments of $\cE$ can reveal only decohered branches to the observer (and cannot reveal their superposition). Such evidence will suggest `quantum jump' from superposition to a single outcome.

\begin{figure*}[tb]
\begin{tabular}{l}
\includegraphics[width=6.3in]{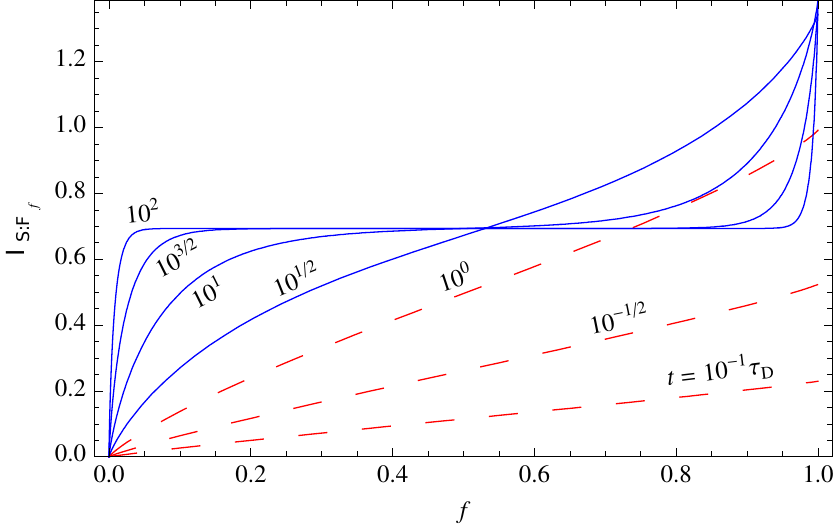}\\
\end{tabular}
\caption{The quantum mutual information $I(\cS : \cF_f )$ vs. fragment size $f$ at different elapsed times for an object illuminated by point-source black-body radiation.
 For point-source illumination, individual curves are labeled by the time $t$ in units of the decoherence time $\tau_D$.
For $t \le\tau_D$ (red dashed lines), the information about the system available in the environment is low.  The linearity in $f$ means each piece of the environment contains new, independent information.  For $t>\tau_D$ (blue solid lines), the shape of the partial information plot indicates redundancy; the first few pieces of the environment increase the information, but additional pieces only confirm what is already known (Riedel and Zurek, 2011). 
}
\end{figure*}

\subsection{Environment as a Witness}

Not all environments are good in this role of a witness. Photons excel: They do not interact with the air or with each other, faithfully passing on information. Small fraction of photon environment usually reveals all we need to know. Scattering of sunlight quickly builds up redundancy: a 1$\mu m$ dielectric sphere in a superposition of 1$\mu m$ size increases ${\cal R}_{\delta=0.1}$ by $ \sim 10^8$ every microsecond (Riedel and Zurek, 2010, 2011). A mutual information plot illustrating this case is shown in Fig. 3.

Air is also good in decohering, but its molecules interact, scrambling acquired data. Objects of interest scatter both air and photons, so both acquire information about position, and favor similar localized pointer states. Moreover, only photons are willing to reveal what was that pointer state.

Quantum Darwinism shows why it is so hard to undo decoherence (Zwolak and Zurek, 2013). Plots of mutual information $I(\cS : \cF_f)$ for initially pure $\cS$ and $\cE$ are antisymmetric (see Figs. 2,~3) around $f= \frac 1 2$ and $H_\cS$ \cite{8}. Hence, a counterpoint of the initial quick rise at $f \le f_\delta$ is a quick rise at $f \ge 1 - f_\delta$, as last few subsystems of $\cE$ are included in the fragment $\cF$ that by now contains nearly all $\cE$. 
This is because an initially pure $\cS \cE$ remains pure under unitary evolution, so $H_{\cS \cE}=0$, and $I(\cS : \cF_f)|_{f=1}$ must reach $2 H_{\cS}$. Thus, a measurement on {\it all} of $\cS \cE$ could confirm its purity in spite of decoherence caused by $\cE \backslash \cF$ for all $f \le 1- f_\delta$. 
However, to verify this one has to intercept and measure all of $\cS\cE$ in a way that reveals pure state $\ket {\Upsilon_{\cS\cE}}$, Eq. (1). Other measurements destroy phase 
information. So, undoing decoherence is in principle possible, but the required resources and foresight preclude it.

In quantum Darwinism decohering environment acts as an amplifier,
inducing branch structure of $\ket {\Upsilon_{\cS\cE}}$ distinct from typical states in the Hilbert space of $\cS\cE$: $I(\cS : \cF_f)$ of a random state is given by the green plot in Fig. 2, with no plateau or redundancy. 
Antisymmetry (Blume-Kohout and Zurek, 2005) means that $I(\cS : \cF_f)$ `jumps' at $f= \frac 1 2$ to $2 H_{\cS}$. 

Environments that decohere $\cS$, but scramble information because of interactions between  subsystems (e.g., air) eventually approach such random states. Quantum Darwinism is possible only when information about $\cS$ is preserved in fragments of $\cE$, so that it can be recovered by observers. There is no need for perfection: Partially mixed environments or imperfect measurements correspond to noisy communication channels -- their capacity is depleted, but we can still get the message (Zwolak et al., 2009; 2020).

Quantum Darwinism settles the issue of the origin of classical reality by accounting for all of the operational symptoms objective existence in a quantum Universe: A single quantum state cannot be found out through a direct
measurement. However, pointer states usually leave multiple records in the environment. Observers can use these
records to find out the (pointer) state of the system of interest. Observers can afford to destroy photons while
reading the evidence -- the existence of multiple copies implies that many can access
the information about the system indirectly and independently, and that they will all agree about
the outcome.  This is how objective existence arises in our quantum world.

There has been significant progress in the study of the acquisition and dissemination of the information by the environments and the consequences of quantum Darwinism. We point interested readers to recent overview (Zurek, 2018) as well as to experimental efforts (see Unden et al. (2019) and references therein), as well as to a popular account (Ball, 2018) as well as to the recent more technical advance (Touil, 2021).

\section{Discussion}

Our Universe is quantum, as experimental confirmations of quantum superpositions on increasingly large scales and as studies entanglement demonstrate. Yet, the world we inhabit appears classical to us. It seems devoid of superpositions or entanglement. 

Emergence of classicality from the quantum substrate was a long-standing mystery. Decoherence accounts for the environment - induced superselection. Einselection leads to preferred pointer states. It posits that objects acquire effective classicality because they are in effect monitored by their environments that destroy superpositions of selected quasi-classical pointer states that are immune to decoherence, as they do not entangle with the environment. 

Quantum Darwinism extends the role of the environment: It is motivated by the insight that we acquire our data about the Universe indirectly, by intercepting fractions of the environment that have caused decoherence. Thus, only information reproduced, by decoherence, in many copies, can be regarded as objective:  It can be accessed by many who will agree about what they perceive, and who can re-confirm their data by measuring additional environment fragments. 

Decohering environment destabilizes all superpositions -- it preserves intact only the einselected pointer states. Einselection is focused on the persistence of states in spite of the environment. It poses and answers a different question than quantum Darwinism which is focused on how the information about the objects we perceive reaches our senses, and, ultimately, our consciousness. However, the answer to both questions revolves around pointer states -- the states that are einselected are also the only states that can be imprinted in many copies on the environment.

Quantum Darwinism recognizes environment's role as a medium, as a communication channel through which information reaches observers\footnote{Similar ideas (albeit with a somewhat different emphasis) are pursued by others, largely within the quantum information community. Recent discussions of the variations on the theme of quantum Darwinism including relevant references can be found in e.g. in Korbicz, (2000), or in Le and Olaya-Castro (2000).}. The ability to obtain information about systems of interest indirectly, by eavesdropping on their environments (rather than through direct measurement) safeguards their pointer states from the disturbance that would be otherwise inflicted by the acquisition of information. Pointer states are selected by the interaction with the environment for their immunity to decoherence. The information about them is inscribed, in multiple copies, on the subsystems of that environment. Only the data about the pointer states is then readily available to observers. 

This is a tradeoff: Observers can acquire information that has predictive value (as pointer states persist in spite of decoherence) but they lose the choice of what to measure -- only the information about the observables selected by the interaction with the environment is within reach (Ollivier et al., 2004; 2005; Zurek, 2018). 

This dual role of the environment (its ``censoring'' of the Hilbert space and its broadcasting of the information about the einselected states) supports and extends the {\it existential interpretation}. Persistence is a precondition of existence. The original focus (Zurek, 1993) of the existential interpretation was preservation of pointer states in presence of decoherence. Zeh (1999) acknowledged the role of the existential interpretation at the time when the emphasis was shifting from einselection to the quantum Darwinian, ``environment as a witness'' point of view. 

Quantum Darwinism adds a new and crucial element to the existential interpretation: It explains how pointer states can be found out without getting disrupted. Thus, the concern about the fragility of quantum states can be -- for macroscopic systems that decohere and that comprise our everyday reality -- put aside. We can find out about them without disrupting their (pointer) states. 

Crucial advance in appreciating the role of the environment as a witness came when the scattering of radiation by a nonlocal ``Schr\"odinger cat'' superposition of dielectric sphere was shown (Riedel and Zurek, 2010; 2011) to produce copious redundancy of the information about its location (see Fig. 3). This is an accurate model of what actually happens: We gain most of our information about macroscopic objects in our Universe by detecting a small fractions of photons they scatter or emit. This is how we find out what exists, by employing the environment as both a witness to the ``crime'' of decohering superpositions and as a guilty party (after all, it is responsible for decoherence in the first place).

Quantum Darwinism is increasingly recognized as key to emergence of the familiar classical reality from within our quantum Universe. Its implications are independent of the interpretational stance, although it does rely on the universal applicability of quantum theory (see, however, Baldij\~ao et al., 2020). It is clearly compatible with the Everett's relative states. It is also compatible with a non-dogmatic reading of Bohr's views, as is decoherence and einselection, as discussed by Camilleri and Schlosshauer (2015).  

Quantum Darwinism answers the key question at the core of the interpretational discussions: How is it possible that fragile quantum states can give rise to robust reality that appears to be objective, undisturbed by the acquisition of information about it by observers? The answer is straightforward -- we use the very same environment that is responsible for decoherence and hence for einselection as a communication channel. Thus, we find out relevant information without interacting directly with the objects of interest.

This millennium has witnessed a significant progress on the interpretational questions that were posed (or at least should have been posed) at the advent of quantum physics. These advances -- decoherence, einselection of the pointer states, and quantum Darwinism -- resolve issues that seemed either impossible to address or remained unnoticed only a few decades ago. These answers are not ``interpretations''. They rely on quantum theory {\it per se}, and do not call on any additional ingredients apart from these needed to pose the questions in the first place. Interpretations can be `adjusted' to accommodate these advances, but it is the  interpretations (Copenhagen or Everett's) that have to adjust: Quantum theory is the basis of decoherence, einselection of the pointer states, and quantum Darwinism, and that basis is non-negotiable -- it does not need to be and cannot be adjusted.

Will -- in view of these advances -- the measurement problem be regarded as ``solved''? I hope it might be (or, at the very least, such a truly significant advance might be `widely acknowledged') but -- to be honest -- I am far from certain. I sense that a large fraction of these who are aware of the measurement problem are also rooting for it to remain a problem. 



Thus, any advance in this area (like decoherence, einselection, or quantum Darwinism) is more often than not greeted with the disappointed ``...but this is just quantum theory...''. 

That is hard to argue with -- it is quantum theory. Indeed, there are authorities with well-deserved stellar reputations whose reaction to the above advances is (at least approximately) ``...but this is just quantum theory...''. 

There are also others who take a more constructive approach and suggest ways of modifying quantum theory that would -- in their opinion -- make it more palatable (Adler, 2003; 2004; Leggett, 1980; 2002; Penrose, 1986; 1989; 1996; `t Hooft, 2014; Weinberg, 1989; 2012). 
The explored possibilities range from a variety of probabilistic approaches (see e.g. references in Baldij\~ao et al., 2020) to superdeterminism (see e.g. Hossenfelder, 2020). Some of these proposals may lead to experimental tests. So far, there is however no experimental evidence for anything other than quantum theory. 

There are of course also physicists   
who acknowledge the importance of the advances discussed here, and many have contributed to both theory and experimental tests of decoherence, einselection, and quantum Darwinism. 

Nevertheless, the question can be posed: Will quantum mechanics be replaced by a more palatable theory...? Personally, I would be delighted if there was a testable alternative, or if it turned out that quantum theory can be deduced from a more fundamental set of simple principles, or if a chink in its armor was found by an experiment, a chink that would hint about how to arrive at such a deeper theory. I applaud efforts to look beyond quantum theory, `as we know it'. 

However, I also strongly suspect that such search is futile, at least in the sense that -- even if quantum theory were to be supplanted by a still deeper theory -- quantum weirdness will not go away as a result: After all, it was the experiment that forced us to accept the inevitability quantum weirdness, and the principle of superposition or entanglement will not go away. I admit that the resistance of general relativity to quantization is intriguing, and heroic attempts to test superpositions of larger and heavier objects (by Arndt, Asplemeyer, Bouwmeester, Zeilinger, and others) are well worth the effort, but I also admit some of my enthusiasm may well be self-serving, based on the conviction that these and other tests of quantum mechanics will continue to reveal more about decoherence and einselection -- that is, the real origin of our everyday classical reality. 

I strongly suspect that the ultimate message of quantum theory is that the separation between what exists and what is known to exist -- between the epistemic and the ontic -- must be abolished. It is hard to argue with ``wanting more'', but wanting more need not imply replacing or abolishing quantum theory. 

Quantum Darwinism, einselection, pointer states, decoherence, etc., are also a result of ``wanting more''. Better understanding of quantum physics and of its relation to the classical world we inhabit led to insights that deepened our appreciation of what quantum theory is about. Almost a century ago, when quantum mechanics was still in its infancy, dissolution of the strict distinction between what exists and what is known, between ontic and epistemic -- characteristic feature of the classical Newtonian physics -- was difficult to envision. Moreover, in the absence of the theory of information -- it was impossible to quantify. 

Surprisingly, as the nascent quantum theory matured, quantum awkwardness did not go away. The lesson of the developments that started with decoherence, and led, through einselection and pointer states to quantum Darwinism and related insights, is that quantum states are epiontic. It now appears that the distinction between what is and what is known about what exists -- something we take for granted in our everyday world -- is not there at the fundamental level: It emerges only as a consequence of the separation between the systems of interest (e.g., macroscopic bodies such as planets, billiard balls, and sometimes perhaps even buckyballs) and the carriers of information about them (such as photons). 


\section{Acknowledgment} 

I have benefited from discussions with Robin Blume-Kohout, Sebastian Deffner, Davide Girolami, Jonathan Halliwell, James Hartle, Pawe\l  \ 
Horodecki, Ryszard Horodecki, Claus Kiefer, Harold Ollivier, Juan Pablo Paz, David Poulin, Hai-Tao Quan, Jess Riedel, Akram Touil, Bin Yan, Michael Zwolak, and, of course, Dieter Zeh. This research was supported by DoE through the LDRD program at Los Alamos, and in part by the Foundational Questions Institute’s Consciousness in the Physical World program, in partnership with the Fetzer Franklin Fund.

\end{document}